\documentclass[aps,preprint]{revtex4}
\usepackage{amssymb}
\usepackage{amsmath}

\input{tcilatex}
\begin{document}

\title{The vortex core excitation spectrum in gapped topological d-wave
superconductors. }
\author{B. Rosenstein$^{1,2}$, I. Shapiro$^{3}$, B. Ya. Shapiro$^{3}$}
\affiliation{$^{1}$Department of Electrophysics, National Chiao Tung University, Hsinchu,
Taiwan, R.O.C. }
\affiliation{$^{2}$Physics Department, Ariel University of Samaria, Ariel 40700, Israel\\
$^{3}$Department of Physics, Institute of Superconductivity, Bar-Ilan
University, Ramat-Gan 52900, Israel}
\keywords{p-wave superconductor, vortex core excitations, Majorana states}
\pacs{PACS: 74.20.Fg, 74.20.Rp, 74.25.Ha}

\begin{abstract}
There are indications that some high $T_{c}$ unconventional superconductors
have a "complex" d-wave order parameter $d_{x^{2}-y^{2}}+i\alpha d_{xy}$
(with an admixture of s-wave) leading to nonzero energy gap $\Delta $. Since
the coherence length is short and the Fermi energy is relatively small the
quasiclassical approach is inapplicable and the more complicated
Bogoliubov-deGennes equations should be used to investigate the excitation
spectrum of such a material in a magneric field. It turns out that equations
for the chiral d-wave superconductor, $\alpha =\pm 1$ simplify considerably
and is the basis for any superconductor of that type with a sufficiently
large gap. The spectrum of core excitations of the Abrikosov vortex in an
anisotropic 3D sample exhibits several features. Unlike in conventional and
gapless superconductors the core has a single excitation mode of order $%
\Delta $ for each value of momentum along the field. This has a large impact
on thermal transport and vortex dynamics.
\end{abstract}

\maketitle

\section{Introduction.}

Initiated by the discovery of the cuprates, the search for new materials
exhibiting unconventional superconductivity has become one of the major
branches in condensed matter physics. The symmetry of the order parameter is
tightly related to the symmetry of the crystal lattice and pairing
mechanism. The bulk condensate is described by a generally tensorial \textit{%
complex} order parameter $\Delta \left( k\right) $ exhibiting a great
variety of the broken symmetries ground states. Generally a time reversal
invariant non s-wave pairing results in nodes in $k$ space, namely the
superconductivity is gapless and the complex nature of the order parameter
does not come into play. However, when the time reversal is broken the
excitation spectrum becomes gapped. Generally two\ real (up to a $k$
independent phase) nodal pairings $\Delta _{1}\left( k\right) $ and $\Delta
_{2}\left( k\right) $ are combined into an essentially complex order
parameter\cite{Sigrist,Annett}, $\Delta =\Delta _{1}+i\Delta _{2}$. The
reason is that nodes of both $\Delta _{1}$ and $\Delta _{2}$ appear
typically along non intersecting lines of the momentum space.

The first system with the gap arising from two nodal pairings was liquid $%
He^{3}$ \cite{Leggett} that in the ABM phase even exhibits the extreme case
of that phenomenon, the so-called chiral superconductivity, characterized by
presence of electron-hole symmetry and absence of both the time-reversal and
spin-rotation symmetry. In this case $\Delta _{1}=p_{x}$ and $\Delta
_{2}=p_{y}$, so that $\Delta =p_{x}+ip_{y}$ with "equal strength" of the two
components rotated by the $\pi /2$ phase "exotic" spin-triplet states.
Charged spin-triplet superfluids, like \cite{Maeno}$Sr_{2}RuO_{4}$ and heavy
fermion compounds like $UPt_{3}$ were shown to have similar structure
although it is not yet clear whether they are chiral. The recently
discovered $Cu$-doped topological superconductor $Bi_{2}Se_{3}$ produces an
equivalent chiral pseudospin system on its surface \cite{Ong}. The surface
states in these materials attract much attention these days because they are
recognized as Majorana fermion bound states \cite{Green}.

Less "exotic" spin-singlet gapped unconventional superconductors are $%
d_{x^{2}-y^{2}}$ wave with admixture of $id_{xy}$. Although most popular
high $T_{c}$ materials $YBCO$ and $BiSCCO$ are pure $d_{x^{2}-y^{2}}$ with
probably small admixture\cite{Okawa} of $id_{xy}$. The admixture was invoked
early on by Laughlin \cite{Laughlin} to explain the time reversal broken
symmetry low-temperature phase observed in $Bi_{2}Sr_{2}CaCu_{2}O_{8}$ via
heat transport experiment in external magnetic field \cite{Krishana},\cite%
{Tanaka}. Similar transition was observed in Ni-doped $BiSCCO$ in zero
external magnetic field. Balatsky \cite{Balatsky} pointed out that, in the
presence of magnetic impurities, the $d_{x^{2}-y^{2}}$ superconductor can
exhibit a transition exactly into the $d+id$ state due to a coupling between
the impurity magnetization and the $d_{xy}$ component of the order
parameter. Topologically nontrivial superconducting state of two-dimensional
electron system was discussed by Kopaev \textit{et al.}\cite{Kopaev} in
connection with the problem of pairing with large center-of-mass pair
momentum under predominant repulsive screened Coulomb interaction. Direct
numerical solution of the self-consistency equation exhibits two nearly
degenerate order parameters $d_{x^{2}-y^{2}}$ and $d_{xy}.$ Spontaneous
breaking of the time-reversal symmetry can mix these states and form fully
gapped chiral $d+id$ superconducting states. Recently an STM evidence
appeared \cite{Wei} of significant admixture of $id_{xy}$ in $Ca$ doped $%
YBCO $.

Theoretically a way to accomplish the mixing-of-components scenario, the
lattice can act as a custodial symmetry to ensure degeneracy of different
superconducting instabilities. In such a case the degeneracy is linked to
higher dimensional irreducible representations of the lattice symmetry
group, and a chiral superposition of superconducting states can be
energetically favorable below $T_{c}$. Cuprates (with an exception of $YBCO$%
) are fourfold symmetric. It was pointed out recently\cite{Kiessel} that for
the square lattice and its $C_{4v}$ group, there is no representation for
singlet Cooper pairs that could result in chiral superconductivity. This,
however, changes for hexagonal systems, where the $E_{2}$ representation of
the $C_{6v}$ lattice symmetry group implies the degeneracy of the $%
d_{x^{2}-y^{2}}$ and $d_{xy}$ wave state at the instability level which can
yield a chiral $d+id$ singlet superconductor. Degeneration of the $%
d_{x^{2}-y^{2}}$ and $d_{xy}$ ordered states inherent in doped graphene
monolayer has been recently considered by Nandkishore \textit{et al.} \cite%
{graphene} as a possible origin of a rise of a singlet chiral
superconducting (SC) state with $d+id$ orbital symmetry.

Chiral d- wave superconductivity (Chern number $\pm 2$) has been suggested
in $Na_{x}CoO_{2}$\textperiodcentered $yH_{2}O_{12-15}$, a novel heavy
fermionic compound \cite{Zhang}. Note that all these systems have small
Fermi momentum $k_{F}$ of the order of magnitude of the inverse coherence
length $\xi ^{-1}$.

Magnetic fields in type II superconducting film easily create stable
line-like topological defects, Abrikosov vortices\cite{Kopnin}. In the
simplest vortex the phase of the order parameter rotates by $2\pi $ around
the vortex and each vortex carries a unit of magnetic flux $\Phi _{0}$.
Quasiparticles near the vortex core "feel" the phase wind by creating a set
of discrete low-energy Andreev bound states. For the $s$-wave
superconductors when the vortices are unpinned (freely moving) these states
were comprehensively studied theoretically including the excitations
spectrum \cite{Caroli}, density of states \cite{Gygi}, their role in vortex
viscosity\cite{Kopnin}, and the microwave absorption \cite{Janko}. The low
lying spectra of quasiparticles and hole excitations are equidistant, $%
E_{l}=l\Delta ^{2}/E_{F},$ where the angular momentum $l$ takes on half
integer values. The "minigap" in the low $T_{c}$ $s$-wave superfluids is of
order of $\Delta ^{2}/E_{F}$. Since the Fermi energy $E_{F}>>\Delta $ 
{\LARGE \ }it is equivalent in the clean limit to large values of
dimensionless parameter $k_{F}\xi >>1$. Roughly there are Andreev bound
states below the superconducting threshold.

Free vortices in the chiral $p$-wave superconductors exhibit a remarkable
topological feature of appearance of the zero energy mode in the vortex core 
\cite{Volovik99}. The spectrum of the low energy excitations remains
equidistant, $E_{l}=\left( l-1\right) \Delta ^{2}/E_{F}$, but now $l$ is
integer\cite{Sigrist}. The zero mode represents a condensed matter analog of
the Majorana fermion first noticed in elementary particle physics \cite%
{Wilczek}. While the minigap in the $s$ and $d$-wave superconductors was
detected by STM, in $p$-wave has not yet been observed. The major reason for
that is the \textit{small} \textit{value of the} \textit{minigap}$\ $in the
core spectrum (just $mK$ for $Sr_{2}RuO_{4}$). It was shown theoretically 
\cite{Melnikov09} that in the $s$-wave superconductors pinning by an
inclusion of radius of just $R=0.2\xi -0.5\xi $ changes dramatically the
subgap excitation spectrum: the minigap $\Delta ^{2}/E_{F}$ becomes of the
order of $\Delta $. \ On the other hand in the chiral p-wave superconductors
the spectrum of the core excitations of the charged states for $R=0.1\xi
-0.4\xi $ is less sensitive to the inclusion, but nevertheless pushes the
spectrum up towards $\Delta $, so that they therefore interfere less with
the Majorana state that is topologically protected and cannot be affected by
the inclusion \cite{Rosenstein}.

In a magnetic field the nodal\ $d_{x^{2}-y^{2}}$ superconductor under the
same "low $T_{c}$" assumption $\gamma ^{-1}=2k_{F}\xi >>1$ that allows the
semiclassical approximation of the spectrum of Andreev states of a singly
vortex was calculated by Kopnin\cite{Kopnin2}. He found the spectrum to be
similar to that of the s-wave superconductor despite the nodes. Maki \cite%
{Maki1} extended the work beyond the semiclassical approximation and found
that there is a series of additional extended states along the node
directions. The low-energy states have no counterpart in a vortex of $s$%
-wave superconductors.

As was pointed out later by Maki and coworkers \cite{Maki2} that in high $%
T_{c}$ cuprates the parameter $\gamma $ is of order $1$ and this modifies
significantly the spectrum. It contains just a few Andreev states in
addition to the node extended states. The situation is expected to be
different in d-wave gapped superconductors with $\gamma ^{-1}\sim 1$
discussed above and therefore the nodal states should disappear, while the
number of Andreev states should be small and could not be treated
semiclassically. In fact the system of Bogoliubov deGennes equations for
d-wave superconductors does not separate into a set of equations for each
angular momentum $l$ unlike that of the s-wave and chiral p-wave. Therefore
beyond the semiclassical approximation all the angular momenta mix and one
is forced\cite{Maki2} to truncate the series. We found that the situation is
different for chiral d-wave case: the system does separates into groups of
two harmonics making its solution possible. For a gapped non-chiral
superconductor one can develop the perturbation theory around the chiral
limit. We perform this calculation for large $\gamma $ and extend the work
to 3D anisotropic superconductors not considered in \cite{Maki2}. We compute
the density of states and thermal conductivity along the field direction
(the vortex axis) that can be effectively used\ along with STM \cite{Hess}
and microwave radiation to detect the features of the pairing.\bigskip

\section{\protect\bigskip The BdG Equations for general d-wave superconductor%
}

\subsection{Microscopic definition of the gap operator}

We begin with the nonlocal BdG equations for the Bogoliubov eigenfunction
corresponding to the eigenenergy $E_{n}$,%
\begin{equation}
\left( 
\begin{array}{cc}
\hat{H}_{0} & \widehat{\Delta } \\ 
\widehat{\Delta }^{+} & -\hat{H}_{0}^{\ast }%
\end{array}%
\right) \left( 
\begin{array}{c}
u_{n} \\ 
v_{n}%
\end{array}%
\right) =E_{n}\left( 
\begin{array}{c}
u_{n} \\ 
v_{n}%
\end{array}%
\right) .  \label{BdG}
\end{equation}%
As a single particle Hamiltonian one can take the parabolic dispersion and
the approximation 
\begin{equation}
H_{0}=\frac{1}{2m}\left( \mathbf{p}-\frac{e}{c}\mathbf{A}\right) ^{2}-E_{F}%
\text{,}  \label{H}
\end{equation}%
where $E_{F}$ is the Fermi energy. The gap function in an unconventional
superconductor\cite{Morita} is characterized by the nonlocal "order
parameter" operator, 
\begin{equation}
\widehat{\Delta }g=\dint \Delta \left( \mathbf{r},\mathbf{r}^{\prime
}\right) g\left( \mathbf{r}^{\prime }\right) d\mathbf{r}^{\prime }\text{.}
\label{delta}
\end{equation}%
Within the BCS theory the kernel $\Delta \left( r,r^{\prime }\right) $ is
subject to the self-consistency condition:

\begin{equation}
\Delta \left( \mathbf{r},\mathbf{r}^{\prime }\right) =\frac{V\left( \mathbf{%
r-r}^{\prime }\right) }{2}\dsum\limits_{n}\left[ u_{n}\left( \mathbf{r}%
\right) v_{n}^{\ast }\left( \mathbf{r}^{\prime }\right) +u_{n}\left( \mathbf{%
r}^{\prime }\right) v_{n}^{\ast }\left( \mathbf{r}\right) \right] \tanh
\left( -\frac{E_{n}}{2T}\right) \text{,}  \label{self}
\end{equation}%
where $V\left( r-r^{\prime }\right) $ is the pairing interaction and $k_{B}$
will be set to $1$.

The vector potential $\mathbf{A}$ for a single vortex has, in polar
coordinates, $r,\varphi $, only an azimuthal component $A_{\varphi }\left(
r\right) $ and in the London gauge consists of the singular part $A_{\varphi
}^{s}=hc/2er$ (field of the infinitely thin solenoid) and typically a rather
insignificant regular part that does not carry flux. Neglecting the regular
part\cite{Sigrist}, the singular one can be "compensated" by the
transformation \cite{Kettersen}:%
\begin{equation}
\left( 
\begin{array}{c}
u_{n} \\ 
v_{n}%
\end{array}%
\right) \rightarrow \left( 
\begin{array}{c}
u_{n}e^{-i\varphi /2} \\ 
v_{n}e^{i\varphi /2}%
\end{array}%
\right) \text{, \ }\Delta \left( \mathbf{r},\mathbf{r}^{\prime }\right)
\rightarrow \Delta \left( \mathbf{r},\mathbf{r}^{\prime }\right)
e^{-i\varphi }\text{.}  \label{transf}
\end{equation}%
After this transformation the Hamiltonian $H_{0}$ has the form

\begin{equation}
H_{0}=-\frac{\hbar ^{2}}{2m_{\bot }}\nabla _{\bot }^{2}-\frac{\hbar ^{2}}{%
2m_{z}}\partial _{z}^{2}-E_{F}\text{.}  \label{H0}
\end{equation}%
Introducing the center-of-mass coordinate, $\mathbf{R}=\left( \mathbf{r+r}%
^{\prime }\right) /2,\mathbf{s=r-r}^{\prime }$ one makes a partial Fourier
representation 
\begin{equation}
\Delta \left( \mathbf{R},\mathbf{s}\right) =\frac{1}{\left( 2\pi \right) ^{2}%
}\dint d\mathbf{k}\Delta \left( \mathbf{R},\mathbf{k}\right) e^{i\mathbf{%
k\cdot s}}.  \label{gap}
\end{equation}%
Expanding $\Delta \left( \mathbf{R},\mathbf{k}\right) $ to second order in $%
\mathbf{k}$,

\begin{equation}
\Delta \left( \mathbf{R},\mathbf{k}\right) =\Delta _{x^{2}-y^{2}}\left(
R\right) \left( k_{x}^{2}-k_{y}^{2}\right) +i\Delta _{xy}\left( R\right)
k_{x}k_{y}\text{,}  \label{del(R)}
\end{equation}%
and substituting it into Eq.(\ref{gap}), one obtains, after integration by
parts, a local form \cite{Zhu} of the order parameter operator:

\begin{equation}
\widehat{\Delta }=-\frac{\Delta _{0}}{4k_{F}^{2}}\left( \widehat{L}_{xx}-%
\widehat{L}_{yy}+2i\alpha \widehat{L}_{xy}\right) \text{.}  \label{deltag}
\end{equation}%
Here the same operators are written via derivatives on the mesoscopic scale: 
$\ \ $%
\begin{equation}
\widehat{L}_{ij}\equiv \left\{ \partial _{i},\left\{ \partial _{j},\Theta
\left( R\right) e^{-i\varphi }\right\} \right\} ,  \label{L}
\end{equation}%
for $i=x,y$. Thus the spatial part of the order parameter $\Delta
_{x^{2}-y^{2}}\left( R\right) =\Delta _{0}\Theta \left( R\right) $ is
normalized by the "isotropic" gap parameter $\Delta _{0}$. We assume for
simplicity that the spatial dependence of the two components is the same $%
\Delta _{xy}\left( R\right) =\alpha \Delta _{0}\Theta \left( R\right) $. The
anti-commutator was defined as $\left\{ \widehat{a},\widehat{b}\right\}
\equiv \widehat{a}\widehat{b}+\widehat{b}\widehat{a}.$\bigskip

\subsection{The chiral representation of the order parameter}

Let us introduce the following "chiral" operators, 
\begin{eqnarray}
L_{++} &=&\left\{ \partial _{+},X_{+}\right\} =L_{xx}-L_{yy}+2iL_{xy}
\label{L++} \\
L_{--} &=&\left\{ \partial _{-},Y_{-}\right\} =L_{xx}-L_{yy}-2iL_{xy} \\
L_{+-} &=&L_{xx}+L_{yy}+iL_{xy}
\end{eqnarray}%
where $\partial _{\pm }=\partial _{x}\pm i\partial _{y}$, and%
\begin{equation}
X_{\pm }=\left\{ \partial _{\pm },\Theta e^{\mp i\varphi }\right\} ;\text{ \
\ }Y_{\pm }=\left\{ \partial _{\pm },\Theta e^{\pm i\varphi }\right\} \text{.%
}
\end{equation}%
In terms of these operators one can write

\begin{equation}
L_{xx}=\frac{1}{4}\left( L_{++}+L_{--}+2L_{+-}\right)  \label{Lplus}
\end{equation}

\begin{equation}
L_{yy}=\frac{1}{4}\left( -L_{++}-L_{--}+2L_{+-}\right)  \label{Lminus}
\end{equation}%
\begin{equation}
L_{xx}-L_{yy}=\frac{1}{2}\left( L_{++}+L_{--}\right) ;L_{xy}=-\frac{i}{4}%
\left( L_{++}-L_{--}\right)  \label{Lxy}
\end{equation}%
Substituting Eqs. \ref{Lplus},\ref{Lminus} and Eq. \ref{Lxy} into Eqs.\ref%
{H0},\ref{deltag}, one obtains in this representation:

\begin{equation}
H_{0}=-\frac{\hbar ^{2}}{2m_{\bot }}L_{+}L_{-}-\frac{\hbar ^{2}}{2m_{z}}%
\partial _{z}^{2}-E_{F};  \label{H00}
\end{equation}

\begin{equation}
\widehat{\Delta }=-\frac{\Delta _{0}}{8k_{F}^{2}}\left( L_{++}\left(
1+\alpha )+L_{--}(1-\alpha \right) \right)  \label{Dg}
\end{equation}

\subsection{Polar coordinates}

In polar coordinates,

\begin{eqnarray}
\partial _{x} &=&-\frac{1}{r}\sin \varphi \partial _{\varphi }+\cos \varphi
\partial _{r};  \label{dx} \\
\partial _{y} &=&\frac{1}{r}\cos \varphi \partial _{\varphi }+\sin \varphi
\partial _{r},  \label{dy}
\end{eqnarray}%
the chiral combinations are

\begin{equation}
\partial _{\pm }=e^{\pm i\varphi }\left( \partial _{r}\pm \frac{i}{r}%
\partial _{\varphi }\right) \text{.}  \label{d+-}
\end{equation}%
The diagonal part of the BdG Hamiltonian takes the form

\begin{equation}
H_{0}=-\frac{\hbar ^{2}}{2m_{\bot }}\left( \frac{1}{r}\partial _{r}+\partial
_{r}^{2}+\frac{1}{r^{2}}\partial _{\varphi }^{2}\right) -\frac{\hbar ^{2}}{%
2m_{z}}\partial _{z}^{2}-E_{F}\text{.}  \label{H01}
\end{equation}%
The off diagonal terms in Eqs.\ref{transf} include:

\begin{eqnarray}
L_{++}e^{i\varphi /2}v_{n} &=&e^{3i\varphi /2}Mv_{n};L_{++}^{+}e^{-i\varphi
/2}u_{n}=e^{-3i\varphi /2}M^{+}u_{n};  \label{L_polar} \\
L_{--}e^{i\varphi /2}v_{n} &=&e^{-i\varphi /2}Nv_{n};L_{--}^{+}e^{-i\varphi
/2}=e^{i\varphi /2}N^{+}v_{n},  \notag
\end{eqnarray}%
where

\begin{equation}
M=\Theta ^{\prime \prime }-\frac{\Theta ^{\prime }}{r}+\frac{2\Theta }{r^{2}}%
+4\Theta ^{\prime }\partial _{r}-\frac{4\Theta }{r}\partial _{r}+\frac{4i}{r}%
\Theta ^{\prime }\partial _{\varphi }-\frac{8i\Theta }{r^{2}}\partial
_{\varphi }+4\Theta \partial _{r}^{2}+\frac{8i\Theta }{r}\partial
_{r}\partial _{\varphi }-\frac{4\Theta }{r^{2}}\partial _{\varphi }^{2};
\label{M++}
\end{equation}

\begin{equation}
N=\Theta ^{\prime \prime }+\frac{3\Theta ^{\prime }}{r}-\frac{2\Theta }{r^{2}%
}+4\Theta ^{\prime }\partial _{r}+\frac{4\Theta }{r}\partial _{r}-4i\frac{%
\Theta ^{\prime }}{r}\partial _{\varphi }-\frac{2i\Theta }{r^{2}}\partial
_{\varphi }+4\Theta \partial _{r}^{2}-8i\frac{\Theta }{r}\partial
_{r}\partial _{\varphi }-\frac{4\Theta }{r^{2}}\partial _{\varphi }^{2}\text{%
.}  \label{N++}
\end{equation}

\bigskip

\section{Solution of the BdG equations}

\subsection{\protect\bigskip Decoupling of different angular momenta in
chiral superconductor}

Let us consider the chiral case of $\alpha =1$ in Eq.(\ref{deltag}). The set
of BdG equation in this case is%
\begin{equation}
\left( 
\begin{array}{cc}
\hat{H}_{0} & -\frac{\Delta _{0}}{4p_{F}^{2}}e^{2i\varphi }\widehat{M} \\ 
-\frac{\Delta _{0}}{4p_{F}^{2}}e^{-2i\varphi }\widehat{M}^{+} & -\hat{H}%
_{0}^{\ast }%
\end{array}%
\right) \left( 
\begin{array}{c}
u_{n} \\ 
v_{n}%
\end{array}%
\right) =E_{n}\left( 
\begin{array}{c}
u_{n} \\ 
v_{n}%
\end{array}%
\right) ,  \label{chiral_BdG}
\end{equation}%
Using the translation symmetry in the field direction, the operator $H_{0}$
for component with momentum $k_{z}$ takes the form

\begin{equation}
\hat{H}_{0}=-\frac{\hbar ^{2}}{2m_{\bot }}\left( \frac{1}{r}\partial
_{r}+\partial _{r}^{2}+\frac{1}{r^{2}}\partial _{\varphi }^{2}\right)
-E_{\perp }\text{,}  \label{H0_final}
\end{equation}%
where $E_{\perp }=E_{F}-k_{z}^{2}/2m_{z}$. Despite the fact that there is no
explicit rotational symmetry, it is convenient to use the 2D angular
momentum basis%
\begin{equation}
\left( 
\begin{array}{c}
u \\ 
v%
\end{array}%
\right) =\dsum\limits_{l}e^{il\varphi }\left( 
\begin{array}{c}
u_{l}\left( r\right) \\ 
v_{l}\left( r\right)%
\end{array}%
\right) ;  \label{angular_momentum}
\end{equation}%
which leads to

\begin{eqnarray}
Eu_{l} &=&e^{-il\varphi }\hat{H}_{0}e^{il\varphi }u_{l}-\frac{\Delta _{0}}{%
4k_{F}^{2}}e^{i\left( 2-l\right) \varphi }\widehat{M}e^{i\left( l-2\right)
\varphi }v_{l-2}  \label{Eq} \\
Ev_{l} &=&-e^{-il\varphi }\hat{H}_{0}e^{il\varphi }v_{l}-\frac{\Delta _{0}}{%
4k_{F}^{2}}e^{-i\left( l+2\right) \varphi }\widehat{M}^{+}e^{i\left(
l+2\right) \varphi }u_{l+2}.  \notag
\end{eqnarray}

With the transition to energies in units of $\Delta _{0}$, in particular $%
\varepsilon _{n}=E_{n}/\Delta _{0}$ and distances in units of coherence
length ($\xi =\hbar v_{F}/\Delta _{0}$ in the clean limit), $r\rightarrow
r/\xi $, $k_{z}\rightarrow \xi k_{z}$, $k_{\perp }\rightarrow \xi k_{\perp }$%
, the equations become:%
\begin{eqnarray}
\varepsilon u_{l} &=&-\gamma \left( \partial _{r}^{2}+\frac{1}{r}\partial
_{r}+k_{\perp }^{2}-\frac{l^{2}}{r^{2}}\right) u_{l}-\gamma ^{2}\Pi
_{1}v_{l-2};  \label{Eq_final} \\
\Pi _{1} &=&\left( \Theta ^{\prime \prime }+\frac{4l-9}{r}\Theta ^{\prime }+2%
\frac{2l^{2}-12l+17}{r^{2}}\Theta +4\Theta ^{\prime }\partial _{r}+4\frac{%
2l-5}{r}\Theta \partial _{r}+4\Theta \partial _{r}^{2}\right) \\
\varepsilon v_{l} &=&-\gamma ^{2}\Pi _{2}u_{l+2}+\gamma \left( \partial
_{r}^{2}+\frac{1}{r}\partial _{r}+k_{\perp }^{2}-\frac{l^{2}}{r^{2}}\right)
v_{l},  \notag \\
\Pi _{2} &=&\left( \Theta ^{\prime \prime }-\frac{4l+9}{r}\Theta ^{\prime }+2%
\frac{2l^{2}+12l+17}{r^{2}}\Theta +4\Theta ^{\prime }\partial _{r}-4\frac{%
2l+5}{r}\Theta \partial _{r}+4\Theta \partial _{r}^{2}\right)
\end{eqnarray}%
where only two dimensionless parameters enter. One is $k_{\perp }\xi $ and
the second is 
\begin{equation}
\gamma =\frac{\Delta _{0}}{4E_{F}}=\frac{1}{2k_{F}\xi }.  \label{gamma}
\end{equation}%
One observes that the equations decouple if one chooses in the second
equation angular momentum $l-2$:

\begin{eqnarray}
\varepsilon u_{l} &=&-\gamma \left( \partial _{r}^{2}+\frac{1}{r}\partial
_{r}+k_{\perp }^{2}-\frac{l^{2}}{r^{2}}\right) u_{l}-\gamma ^{2}\Pi
_{3}v_{l-2};  \label{final_eq} \\
\Pi _{3} &=&\left( \Theta ^{\prime \prime }+\frac{4l-9}{r}\Theta ^{\prime }+2%
\frac{2l^{2}-12l+17}{r^{2}}\Theta +4\Theta ^{\prime }\partial _{r}+4\frac{%
2l-5}{r}\Theta \partial _{r}+4\Theta \partial _{r}^{2}\right) \\
\varepsilon v_{l-2} &=&-\gamma ^{2}\Pi _{4}u_{l}+\gamma \left( \partial
_{r}^{2}+\frac{1}{r}\partial _{r}+k_{\perp }^{2}-\frac{\left( l-2\right) ^{2}%
}{r^{2}}\right) v_{l-2}.  \notag \\
\Pi _{4} &=&\left( \Theta ^{\prime \prime }-\frac{4l+1}{r}\Theta ^{\prime }+2%
\frac{2l^{2}+4l+1}{r^{2}}\Theta +4\Theta ^{\prime }\partial _{r}-4\frac{2l+1%
}{r}\Theta \partial _{r}+4\Theta \partial _{r}^{2}\right)
\end{eqnarray}

The profile of the order parameter $\Theta \left( r\right) $ should be
calculated self consistently, however we used a simple dependence $\Theta
\left( r\right) =\tanh \left( r\right) $. This is justified a posteriori by
the local density of states (LDOS) that shows that it is completely
dominated by the continuum states rather than the few Andreev states.

\subsection{Results for the spectrum and charge density for a single vortex}

This was solved numerically with boundary conditions $u_{l}\left( r=0\right)
=v_{l}\left( r=0\right) =0$ and $u_{l}\left( r=L\right) =v_{l}\left(
r=L\right) =0$, where $L$ is the radius of the cylindrical sample (disc when
the width of the sample is finite $-L_{z}/2<z<L_{z}/2$). Pinned vortices are
discussed in the next section. The Andreev bound state is found for small $%
\gamma $ only for $l=1$. The energy as function of $k_{z}$ for $\gamma =0.38$
is presented in Fig.1.

\bigskip

\bigskip \FRAME{ftbpFU}{6.1967in}{4.7513in}{0pt}{\Qcb{A single Andreev state
energy as function of the momentum along the magnetic field direction. The
value of the only parameter characterizing the system is $\protect\gamma %
=0.38$. The dispersion relation is nearly linear up to a threshold at $%
\Delta $.}}{\Qlb{Fig1}}{Fig1}{\special{language "Scientific Word";type
"GRAPHIC";maintain-aspect-ratio TRUE;display "USEDEF";valid_file "F";width
6.1967in;height 4.7513in;depth 0pt;original-width 8.2in;original-height
6.2798in;cropleft "0";croptop "1";cropright "1";cropbottom "0";filename
'C:/Documents and Settings/Boris Shapiro/Desktop/pic
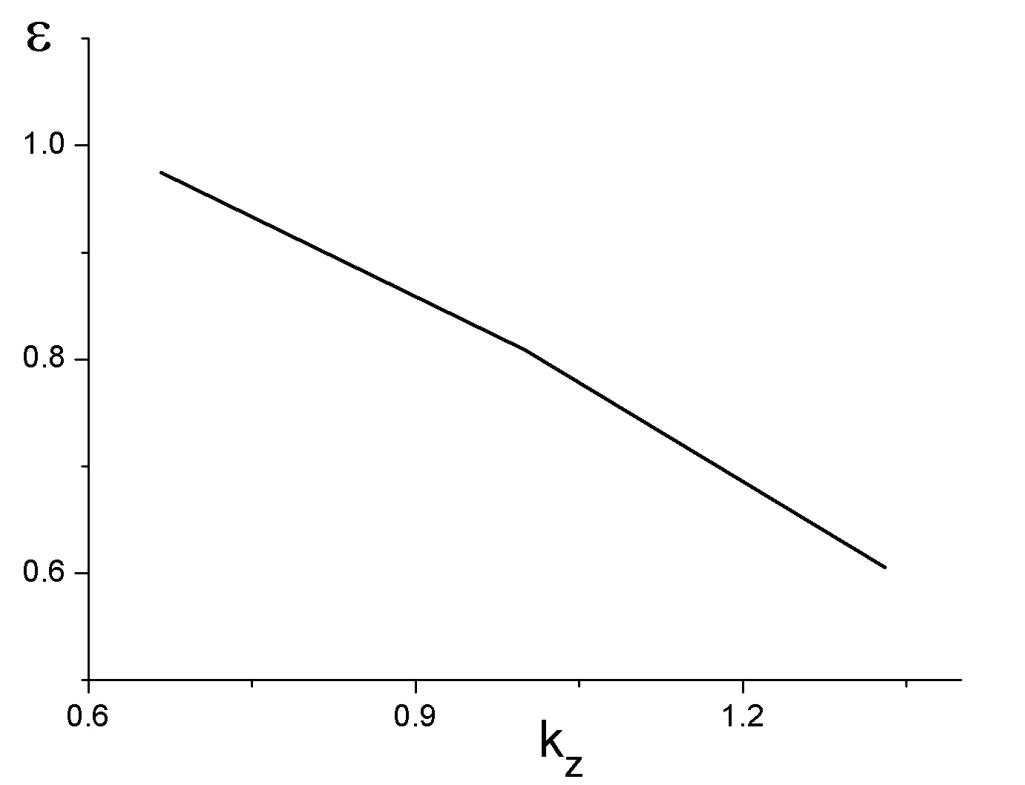';file-properties "XNPEU";}}

\bigskip

All the other angular momentum channels have just the continuum above the
superconducting threshold. Moreover it appears only for 
\begin{equation}
k_{z}<k_{z}^{\max }=2/\gamma  \label{kmax}
\end{equation}%
since $E_{\perp }$ should be positive. This is a direct consequence of small 
$\gamma $. For large $\gamma $ the spectrum of Andreev states is
semiclassical as discussed in the Introduction. The "minigap" therefore is
very large and it is difficult to excite the core states. The Andreev states
(one for each available value of $k_{z}$) are localized near the core, see
the radial density 
\begin{equation}
\rho \left( r\right) =2\pi r\left( \left\vert u_{1}\left( r\right)
\right\vert ^{2}+\left\vert v_{1}\left( r\right) \right\vert ^{2}\right)
\label{rho}
\end{equation}%
in Fig.2. \ It extends several coherence lengths inside the
superconductor\bigskip \bigskip

\bigskip \FRAME{ftbpFU}{6.5388in}{4.7488in}{0pt}{\Qcb{The radial density $%
\protect\rho \left( r\right) $, Eq.(\protect\ref{rho}), of the core state as
function of the distance from the vortex center. for $\protect\gamma %
=0.38,k_{z}=0.88/\protect\xi $}}{\Qlb{Fig2}}{2.jpg}{\special{language
"Scientific Word";type "GRAPHIC";maintain-aspect-ratio TRUE;display
"USEDEF";valid_file "F";width 6.5388in;height 4.7488in;depth
0pt;original-width 8.6401in;original-height 6.264in;cropleft "0";croptop
"1";cropright "1";cropbottom "0";filename 'C:/Documents and Settings/Boris
Shapiro/Desktop/pic 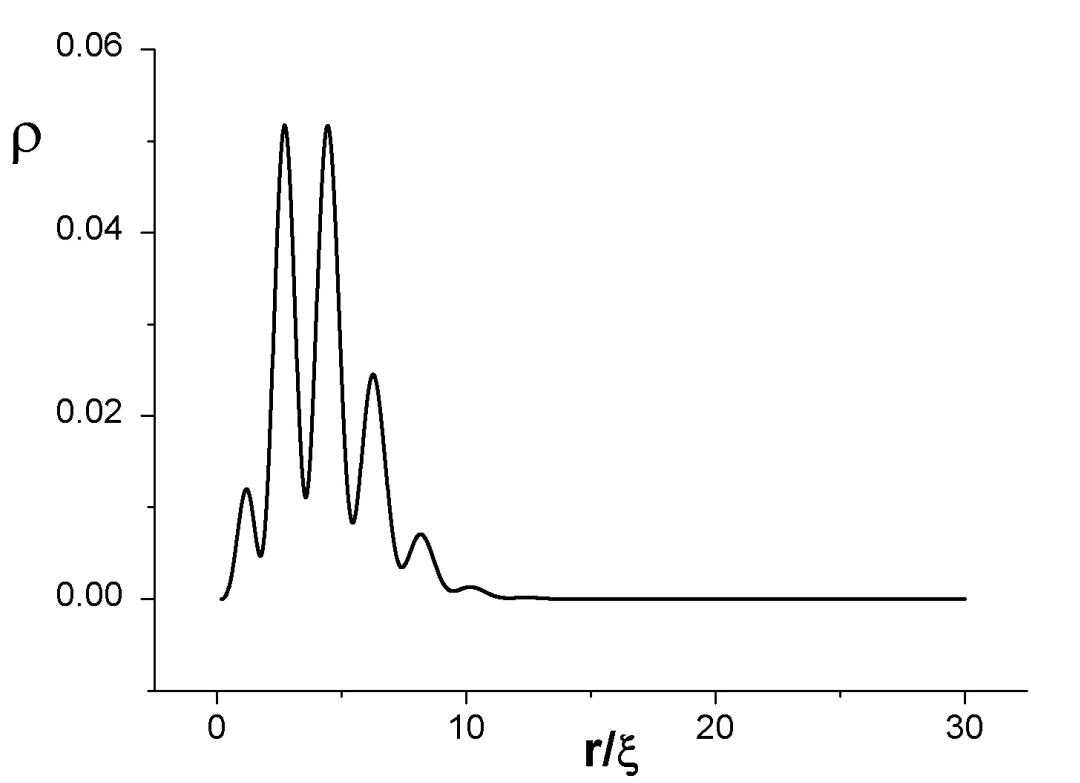';file-properties "XNPEU";}}

\bigskip

The core states come mostly from the particle rather than the hole sector
for the given direction of the magnetic field, so there is a sizable charge
density in the hole. The contributions from the continuum above the
threshold are expected to compensate each other since they originate both
from electron and hole excitations. The density of states reads: 
\begin{eqnarray}
N\left( r,\varepsilon \right) &=&\sum_{n}\left\{ \left\vert u_{n}\left(
r\right) \right\vert ^{2}\delta \left( \varepsilon -\varepsilon _{n}\right)
+\left\vert v_{n}\left( r\right) \right\vert ^{2}\delta \left( \varepsilon
+\varepsilon _{n}\right) \right\} =  \label{LDOS} \\
&=&\frac{\left\vert u_{1}\left( r\right) \right\vert ^{2}}{\sqrt{\varepsilon
-\varepsilon _{1}\left( 0\right) }}+\sum_{cont}\left\{ \frac{\left\vert
u_{n}\left( r\right) \right\vert ^{2}}{\sqrt{\varepsilon -\varepsilon
_{n}\left( 0\right) }}+\frac{\left\vert v_{n}\left( r\right) \right\vert ^{2}%
}{\sqrt{\varepsilon +\varepsilon _{n}\left( 0\right) }}\right\} .  \notag
\end{eqnarray}

\subsection{Generalization to the nonchiral case and to pinned vortices}

Generally the $d$ - wave order parameter, Eq.(\ref{Dg}), deviates from
"perfect chirality", $\alpha =\pm 1$, Small deviations from the positive
chirality can be parametrized by $\zeta $, $\alpha =1+2\zeta $:

\begin{equation}
\widehat{\Delta }=-\frac{\Delta _{0}}{4k_{F}^{2}}L_{++}-\zeta \frac{\Delta
_{0}}{4k_{F}^{2}}\left( L_{++}-L_{--}\right) \text{.}
\end{equation}%
The correction therefore can be written using Eq.(\ref{Lxy}) as%
\begin{equation}
V=-i\zeta \frac{\Delta _{0}}{k_{F}^{2}}L_{xy}\text{.}
\end{equation}%
The shift of the Andreev eigenstates energy occurs only in second order in $%
\zeta $ since%
\begin{equation}
\Delta \varepsilon _{1}=-i\zeta \frac{\Delta _{0}}{k_{F}^{2}}%
\int_{r=0}^{L}r\int_{\varphi }\left\langle l=1\left\vert L_{++}\right\vert
l=1\right\rangle +cc=0\text{.}  \label{zero_shift}
\end{equation}%
The Andreev state goes lower therefore only when the energy gap $\Delta
=\Delta _{0}\left( 1-2\zeta \right) $ is significantly (first order in $%
\zeta $) smaller than $\Delta _{0}$. We expect that for yet lower $\alpha $
the spectrum at small $\gamma $ will still consist of just one Andreev state
in the $l=1$ channel.

Till now the boundary condition at the center of the vortex was free. One
can describe a pinned vortex by a different boundary condition, $u_{l}\left(
r=R\right) =v_{l}\left( r=R\right) =0$, where $R$ is the radius of the
pinning area of order $\xi $ (assumed for simplicity dielectric). The
results do not change appreciably until the radius exceeds $\xi $. \ 

\section{Thermal transport}

\subsection{Kopnin-Landauer formula}

Thermal conductivity is an effective tool to demonstrate the minigap due to
activated behavior of the electron contribution. To calculate the
quasiparticle contribution to thermal conductivity along the vortex cores
when the upper side of the vortex line is held at temperature $T_{1}$ and
the lower side at temperature $T_{2}$, we use a general ballistic (width of
the film $L_{z}$ smaller than the mean free path) Kopnin-Landauer formula%
\cite{Kopnin03}. The heat current at temperature lower than the threshold to
the continuum of states is carried mainly by the bound core states. For a
single vortex in a sufficiently thick sample the variable $k_{z}$ can be
considered as a continuous one and the thermal current can be written as

\begin{eqnarray}
\text{\ }I\left( T\right) &=&\int_{0}^{k_{z}^{\max }}\frac{dk_{z}}{2\pi
\hbar }\left\vert \frac{dE_{1}\left( \gamma ,\xi k_{z}\right) }{dk_{z}}%
\right\vert \frac{E_{1}\left( \gamma ,\xi k_{z}\right) }{1+\exp \left(
E_{1}\left( \gamma ,\xi k_{z}\right) /T\right) }=  \label{I} \\
&=&\frac{\Delta _{0}^{2}}{2\pi \hbar }\int_{0}^{k_{z}^{\max }\xi }d%
\widetilde{k}_{z}\frac{d\varepsilon _{1}\left( \gamma ,\widetilde{k}%
_{z}\right) }{d\widetilde{k}_{z}}\frac{\varepsilon _{1}\left( \gamma ,%
\widetilde{k}_{z}\right) }{1+\exp \left( \varepsilon _{1}\left( \gamma ,%
\widetilde{k}_{z}\right) /t\right) },
\end{eqnarray}%
where $t=T/\Delta _{0}$ and $\widetilde{k}_{z}=\xi k_{z}$. Changing
variables one obtains 
\begin{equation}
\text{\ }I\left( T\right) =\frac{\Delta _{0}^{2}}{2\pi \hbar }%
\int_{1}^{2/\gamma }\frac{\varepsilon d\varepsilon }{1+\exp \left(
\varepsilon /t\right) }=\frac{T^{2}}{2\pi \hbar }\left[ \Pi \left( \frac{%
\Delta _{0}}{T}\right) -\Pi \left( \frac{8E_{F}}{T}\right) \right] .
\label{I(T)_final}
\end{equation}%
Here the lower limit of integration is the energy for $k_{z}=0$ and the
indefinite integral is 
\begin{equation}
\Pi \left( \varepsilon \right) =\int \frac{\varepsilon d\varepsilon }{1+\exp
\left( \varepsilon \right) }=-\varepsilon ^{2}/2+\varepsilon \log \left(
1+e^{\varepsilon }\right) +Li_{2}\left( -e^{\varepsilon }\right) \text{,}
\label{Pi}
\end{equation}%
where $Li$ is the polylog function. For small temperature differences the
linear response can be used,

\begin{equation}
\frac{dI}{dT}=\frac{T}{\pi \hbar }\left[ \Pi \left( \frac{\Delta _{0}}{T}%
\right) +\frac{\Delta _{0}^{2}/2T^{2}}{1+\exp \left( \Delta _{0}/T\right) }%
-\Pi \left( \frac{8E_{F}}{T}\right) -\frac{32E_{F}^{2}/T^{2}}{1+\exp \left(
8E_{F}/T\right) }\right] \text{.}  \label{minigapth}
\end{equation}

\bigskip

\bigskip \FRAME{ftbpFU}{6.5944in}{4.7488in}{0pt}{\Qcb{The temperature
difference between the bottom and the top contacts leads to energy flow
carried at low temperatures by the core states. Dimensionless heat
conductance\ $\frac{\hbar }{\Delta }\frac{dI}{dT}$ of a single vortex given
in Eq.(\protect\ref{minigapth}) as function of temperature in units of $%
\Delta $.}}{\Qlb{Fig3}}{3.jpg}{\special{language "Scientific Word";type
"GRAPHIC";maintain-aspect-ratio TRUE;display "USEDEF";valid_file "F";width
6.5944in;height 4.7488in;depth 0pt;original-width 8.7438in;original-height
6.2881in;cropleft "0";croptop "1";cropright "1";cropbottom "0";filename
'C:/Documents and Settings/Boris Shapiro/Desktop/pic
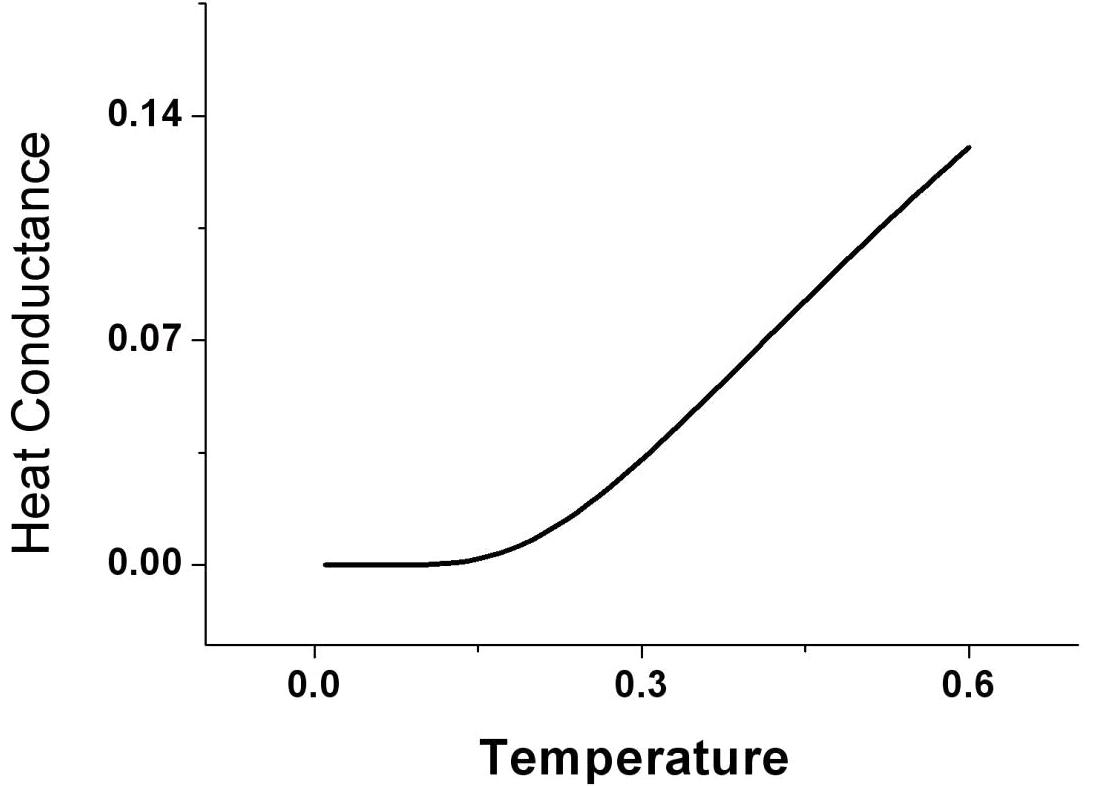';file-properties "XNPEU";}}

\bigskip

\bigskip\ 

In Fig.3 the heat conductance of a single vortex line is given as function
of the inverse temperature (in units of $\Delta ^{-1}$) demonstrating the
activated behaviour of the electron contribution.

\bigskip

\section{\protect\bigskip Conclusion and discussion}

To summarize, in the gapped $d_{x^{2}-y^{2}}+i\alpha d_{xy}$ unconventional
(in the sense of $E_{F}$ being of the same order of magnitude as $\Delta $)
superconductor in a magnetic field the spectrum consists of a single
excitation mode for any value of momentum $k_{z}$ along the field per
Abrikosov vortex. Therefore each vortex core can be viewed as a nano-size
normal "quantum wire" inside the superconducting material. The dispersion in
this one-dimensional metallic system is roughly linear as a function of $%
k_{z}$, see Fig.1. Unlike the extended Andreev states found in a better
studied theoretically case of \ the nodal $d_{x^{2}-y^{2}}$ case\cite{Maki1}%
, the core states are well localized, see Fig.2.\ 

Due to the exceptionally large "minigap" it is difficult to excite the 1D
quantum wire mechanically, thermally or electromagnetically. This has a
large impact on the thermal transport along the field direction and vortex
dynamics in the direction perpendicular to the field. At temperatures lower
that $\Delta /2$ the viscosity should be very large, while the thermal
transport has an activated nature see Fig. 3. In particular case the
critical current in such superconductors would be greatly enhanced at these
temperatures. A magnetic field $B>>H_{c1}$ creates $SB/\Phi _{0}$ vortices
over area $S$, so that heat conductivity is $\kappa =\frac{L_{z}B}{\Phi _{0}}%
\frac{dI}{dT}$, where $L_{z}$ is the sample width. For $B=5T$ (between $%
H_{c1}$ and $H_{c2}$ for cuprates), with $L_{z}=70nm$ at $T/\Delta =0.2$ one
obtains a thermal conductivity of order $\kappa \sim 10^{2}W/Km$.

Technically, since the coherence length is short and the Fermi energy is
relatively small, the quasiclassical approach is inapplicable and more
complicated Bogoliubov-deGennes equations were used. The approach simplifies
for the chiral $d$-wave superconductor, $\alpha =\pm 1,$ due to the
decoupling of sectors in the BdG equation with different angular momentum $l$%
. Despite the fact that the angular momentum is not a conserved quantum
number, it can be used to label Andreev bound states. It turns out that such
states exist only in the $l=1$ channel. The approach can be generalized for
any superconductor of that type with a sufficiently large gap. A natural and
rather direct method to look for evidence for chiral d-wave superconductors
is microwave absorption by the vortex core states in a magnetic field. In
this case the absorption depends strongly on the polarization of the
incident wave \cite{micro}.

Vortices are typically organized in the hexagonal vortex lattice that leads
to the creation of a narrow band due to small overlaps of the Andreev states
belonging to the neighbouring vortex cores. Periodicity however is not
expected to play a role in thermal conductivity along the "nanowires" that
are well separated.

\bigskip

\textit{Acknowledgements.} B.Ya.S. and I.S. acknowledge support from the
Israel Scientific Foundation.

\end{document}